\documentclass[letterpaper]{article} 
\usepackage[submission]{aaai25}  
\usepackage{times}  
\usepackage{helvet}  
\usepackage{courier}  
\usepackage[hyphens]{url}  
\usepackage{graphicx} 
\usepackage{natbib}  
\usepackage{caption} 
\usepackage{amsmath} 
\usepackage{booktabs}
\usepackage{multirow}
\usepackage{array}
\usepackage{adjustbox}
\usepackage{amssymb} 
\usepackage{pifont}
\usepackage{graphicx} 
\usepackage{subcaption}
\newcommand{\cmark}{\ding{51}}%
\newcommand{\xmark}{\ding{55}}%

\usepackage{algorithm}
\usepackage{algorithmic}

\usepackage{newfloat}
\usepackage{listings}
\DeclareCaptionStyle{ruled}{labelfont=normalfont,labelsep=colon,strut=off} 
\floatstyle{ruled}
\newfloat{listing}{tb}{lst}{}
\floatname{listing}{Listing}
\title{DuoLift-GAN:Reconstructing CT from Single-view and Biplanar X-Rays \\ with Generative Adversarial Networks}
\author {
    Zhaoxi Zhang,
    Yueliang Ying
}
\affiliations {
    University of North Carolina at Chapel Hill, Chapel Hill, NC, 27514, USA
}

\usepackage{bibentry}

\begin{document}

\maketitle

\begin{abstract}

Computed tomography (CT) provides highly detailed three-dimensional (3D) medical images but is costly, time-consuming, and often inaccessible in intraoperative settings~\cite{world2011baseline}. Recent advancements have explored reconstructing 3D chest volumes from sparse 2D X-rays, such as single-view or orthogonal double-view images. However, current models tend to process 2D images in a planar manner, prioritizing visual realism over structural accuracy. In this work, we introduce DuoLift Generative Adversarial Networks (DuoLift-GAN), a novel architecture with dual branches that independently elevate 2D images and their features into 3D representations. These 3D outputs are merged into a unified 3D feature map and decoded into a complete 3D chest volume, enabling richer 3D information capture. We also present a masked loss function that directs reconstruction towards critical anatomical regions, improving structural accuracy and visual quality. This paper demonstrates that DuoLift-GAN significantly enhances reconstruction accuracy while achieving superior visual realism compared to existing methods.

\end{abstract}


\section{Introduction}
Radiography, or X-ray imaging, is a medical imaging procedure that produces 2D projection images of a scanned patient with relatively low radiation doses and fast acquisition speeds. In contrast, computed tomography (CT)\cite{buzug2011computed} is an imaging modality that uses a series of X-rays (typically between 100 and 250) taken at different angles around 360 degrees from which a three-dimensional volume is reconstructed. A key advantage of CT imaging is its ability to provide precise spatial information about anatomical structures, such as the lungs, in 3D space. However, CT procedures are time-consuming and expensive. X-rays, by comparison, are quicker, less costly, and result in significantly less radiation exposure~\cite{world2011baseline}. 2D X-ray images lack the 3D spatial information necessary for medical problems requiring quantitative morphological analysis. Thus, reconstructing a 3D CT volume from orthogonal double-view X-rays or a single-view X-ray could provide approximate 3D information and combine X-ray and CT imaging benefits. Although full 3D precision cannot be expected in such an approach, a significant 3D context may be inferred by learning from data.


The primary challenge in reconstructing a 3D CT image from two orthogonal X-ray images or a single-view X-ray image is the need for more spatial information, particularly along the axis perpendicular to the 2D X-ray image. X-ray measurements capture the attenuation of X-rays as they pass through various tissues along their path. As an inverse problem, CT reconstruction typically requires projections densely sampled over 180 degrees. Fewer projections increase the ambiguity in solving this inverse problem using conventional CT reconstruction methods, such as Filtered Backprojection~\cite{Chetih2015}. However, with recent advancements in deep learning and the availability of large-scale medical image datasets, researchers have developed CT reconstruction models~\cite{ying2019x2ct,kasten2020end,ge2022x,gao20233dsrnet,song2021oral,kyung2023perspective,sun2023ct} that can reconstruct 3D volumes from orthogonal double-view X-rays or even single-view X-ray images by leveraging learned priors from datasets.


Some studies concentrate on dense anatomical structures, such as the knees~\cite{kasten2020end}, cervical vertebrae~\cite{ge2022x}, and spine~\cite{gao20233dsrnet}, which contain bones with high attenuation coefficients, resulting in high contrast in radiographic images. Other research focuses on soft tissue reconstruction, particularly the lung~\cite{ying2019x2ct,kyung2023perspective,sun2023ct}, an air-filled organ with a low attenuation coefficient inside. However, existing methods primarily target the reconstruction of the lung's overall shape and surrounding soft tissues, often neglecting the intricate internal structures, such as the numerous thin, tree-like pulmonary vessels. These delicate structures are relatively smaller than the lung, resulting in subtle and complex features in 2D radiographic images.

To address the challenges in 2D-to-3D reconstruction, we considered a method of lifting 2D images and features to 3D in advance to mitigate the difficulty associated with learning the mapping relationship between different dimensions. So we introduce the Duo Lift Generative Adversarial Networks (DuoLift-GAN), using a masked loss function; our model enhances detail capture by aligning the reconstruction with the target chest volume. By replicating the original 2D images and their corresponding feature maps along a specific axis multiple times, we elevate 2D images and their features into 3D representations that preserve spatial relationships and spatial coherence, resulting in more accurate 3D feature maps and improved reconstruction quality. Additionally, for precise shape and contour reconstruction, we present DuoLift-CNN, which accurately reconstructs larger anatomical structures, such as the lungs, which are suitable for capturing the overall structure rather than fine and thin structures, such as the vessels.


Moreover, we observed an intriguing phenomenon during our evaluation of the reconstruction results: while GANs tend to produce volumes with richer textural details, CNN models outperform them in numerical metrics. To further explore this discrepancy, we conducted an in-depth analysis of the evaluation metrics used for CT reconstruction, focusing on the Structural Similarity Index Measure (SSIM)\cite{wang2004image}, Peak Signal-to-Noise Ratio (PSNR)\cite{hore2010image}, and the Learned Perceptual Image Patch Similarity (LPIPS)\cite{zhang2018perceptual} metric. This trend is also evident in the quantitative results of X2CT\cite{ying2019x2ct}. Moreover, we conduct the anatomical structure level quantitative analysis of the reconstruction quality via the Dice Coefficient (DICE)\cite{dice1945measures} of the lung and vessel segmentation maps. With the broad spectrum of evaluation metrics from pixel level, anatomical structure level, to perceptual level, we conduct a comprehensive quantitative and qualitative analysis of our methods and the state-of-the-art method on one lung CT dataset, the LIDC-IDRI dataset. Our analysis evaluates the reconstruction performance and provides insightful observations for the chest CT reconstruction problem from orthogonal X-rays.


In conclusion, our contributions include:
\begin{enumerate}
  \item  We introduce a novel 2D-to-3D reconstruction architecture with dual branches that independently elevate 2D images and their feature maps into 3D representations. These branches preserve spatial relationships and 3D structure, with their outputs merged into a unified 3D feature map. Additionally, we employ a masked loss to train the generator in conjunction with a discriminator, which improves the accuracy of generated textural details.  
  \item Our models, DuoLift-GAN and DuoLift-CNN, achieve the best performance on benchmarks compared to methods with available code.
  \item We conduct an in-depth evaluation of the reconstruction quality on the anatomical structure level.
  \item We will release our code and pre-trained model weights on GitHub, promoting further research and development.
\end{enumerate}


\section{Related Work}

\subsection{CT reconstruction from single or double view X-rays}

Reconstructing accurate 3D volumes from single-view and orthogonal double-view X-rays is challenging, as significant 3D spatial information must be inferred from limited 2D data. Various algorithms tackle this by incorporating back-projection, such as PerX2CT~\cite{kyung2023perspective}, Xray2CT~\cite{sun2023ct}, and C$^2$RV~\cite{lin2024c2rv}. These models first encode X-rays with 2D convolutional layers and then back-project the 2D feature maps into 3D space, followed by decoding the geometry-aware 3D feature maps via 2D convolutional layers~\cite{kyung2023perspective} or multi-layer perceptrons (MLPs)~\cite{sun2023ct,lin2024c2rv}. Other approaches extract 3D features from 2D X-rays by treating the feature dimension of the 2D feature maps as the depth dimension. For example, X2CT-GAN~\cite{ying2019x2ct}, SdCT-GAN~\cite{cheng2023sdct}, and MFCT-GAN~\cite{jiang2022mfct} encode 2D images with 2D convolutional layers, lift the 2D feature maps by duplicating them along the depth dimension, and then decode the lifted 3D feature maps into the 3D volumes. On the other hand, 3DSP-GAN~\cite{huang20243dsp} directly duplicates frontal and lateral X-rays to transform the 2D images to 3D volumes and then pass the 3D volumes to the reconstruction networks. Our model introduces dual branches that independently elevate 2D images and their features into 3D representations for capturing richer 3D information. These 3D feature maps from the dual branches are merged into a unified 3D feature map.

\subsection{Chest CT Reconstruction from X-rays}
One of the challenges in chest reconstruction is ensuring that the results look authentic, avoiding the appearance of fuzzy artifacts inside the lung. Compared to works focusing on Chest CT reconstruction from multi-view such as Xray2CT~\cite{sun2023ct} and C$^2$RV~\cite{lin2024c2rv}, the artifact issue is more critical for single-view or double-view chest CT reconstruction~\cite{ying2019x2ct, cheng2023sdct, kyung2023perspective}. X2CT-GAN~\cite{ying2019x2ct} addresses this by using a discriminator to enhance textural details. SdCT-GAN~\cite{cheng2023sdct} goes further by feeding both the reconstructed and target volumes and their cropped versions into the Discriminator to incorporate more detailed information. We propose to apply a mask generated from the segmentation of the target chest volume to both the target and reconstructed volumes and then calculate the similarity loss between them to train the Generator and combine it with the Discriminator to reconstruct textural details more accurately.

\subsection{Reconstruction Quality Evaluation}
The two most commonly used metrics for assessing the quantitative results between the target and reconstructed volumes are the Structural Similarity Index (SSIM) and Peak Signal-to-Noise Ratio (PSNR)~\cite{wang2004image,hore2010image,sara2019image,setiadi2021psnr}. SSIM is an image quality metric that assesses degradation by measuring perceived structural changes, while PSNR is a metric that quantifies image or video quality by comparing the maximum potential signal power (original data) to the power of distorted noise. However, based on our experiments and the results from X2CT, these two metrics tend to be biased toward larger structures, such as the contours of the chest and lung and are less sensitive to the quality of textual details within the chest. To address this limitation, we introduce additional metrics. The Learned Perceptual Image Patch Similarity (LPIPS)~\cite{zhang2018perceptual} metric calculates the perceptual similarity between two images and is used in SdCT-GAN to evaluate the chest volume. Furthermore, the Dice Coefficient (DICE), a common metric for assessing the overlap between predicted binary results and ground truth, is widely used to evaluate segmentation and registration models. In our evaluation, we applied segmentation models to obtain anatomical structure-level segmentation masks, such as the lungs and vessels. We evaluate the DICE of the lung and the vessel shapes between the target and reconstructed volume, providing a detailed assessment of the model's performance in capturing these critical anatomical structures.

\begin{figure*}[htp]
    \centering
    \includegraphics[width=1.0\linewidth]{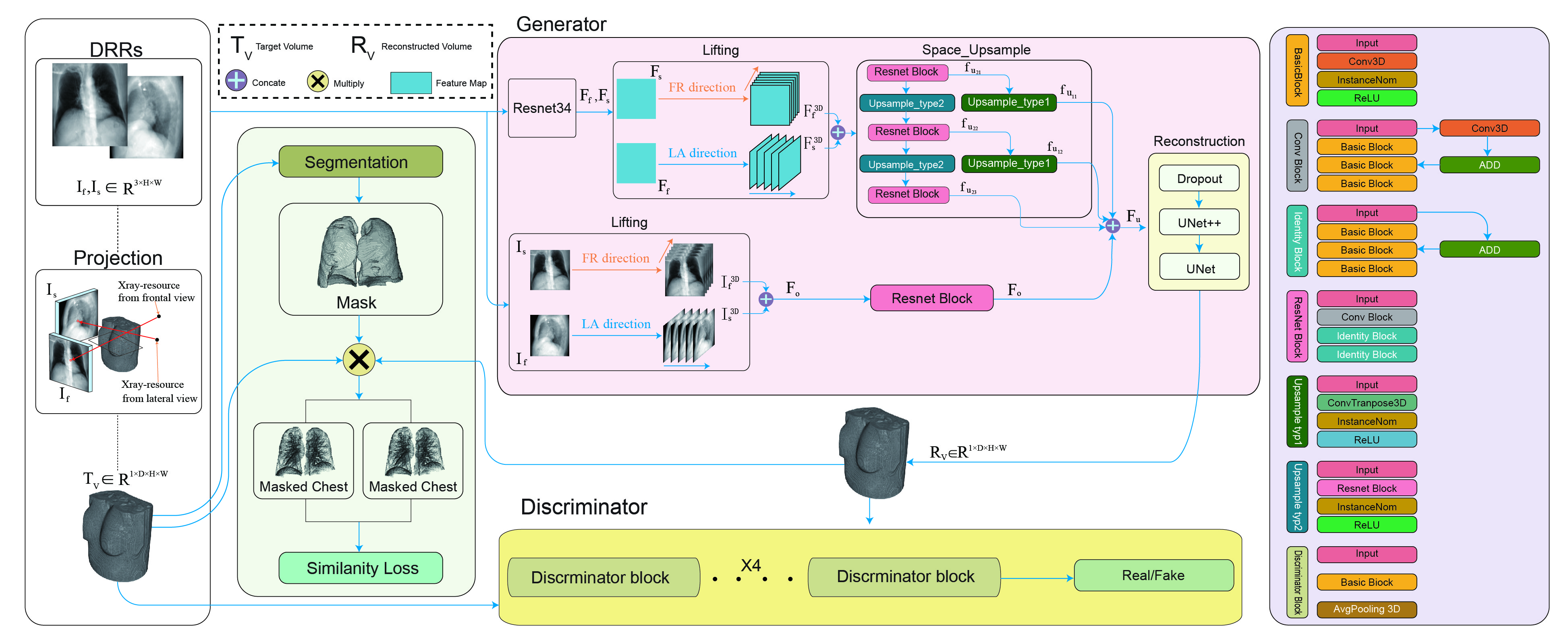}
    \caption{Network architecture of DuoLift-GAN. From left to right, DRRs produced through forward projection on target volume are passed into the Generator to generate the volumetric reconstruction. The generated reconstruction and target volume are then input into the Discriminator.}
    \label{fig:network_archtecture}
\end{figure*}

\section{Method}
Building on the concept of Conditional GAN~\cite{ying2019x2ct}, we propose DuoLift-GAN, a generative model designed for chest volume reconstruction from X-rays. The architecture of DuoLift-GAN, illustrated in Fig.~\ref{fig:network_archtecture}, includes a Generator (G) and a Discriminator (D). We denote the input in the single-view X-ray setting as $I_{f}\in{R}^{H\times{W}}$ which is a coronal chest projection, and the inputs in the orthogonal double-view X-rays setting as $I_{f}$ and $I_{s}$. Given $I_f$ or ${I_f, I_s}$ as the input, our goal is to train a model that can reconstruct the corresponding target chest volume $T_{v}\in{R}^{D^{\prime}\times{H^{\prime}}\times{W^{\prime}}}$.   


\subsection{Generator Architecture}
Our generator employs two branches to extract 3D features from the 2D X-rays. In the first branch, the input projections are passed through a ResNet-34~\cite{he2016deep} 2D feature encoder to extract 2D features, denoted as $F^{2D}_{f,s}\in{R}^{512\times\frac{H}{4}\times\frac{W}{4}}$, from the input images $I_{f,s}$. These 2D features are then lifted into 3D feature maps, $F^{3D}_{f,s}\in{R}^{512\times\frac{D^{\prime}}{4}\times\frac{H^{\prime}}{4}\times\frac{W^{\prime}}{4}}$, by repeating $F^{2D}_{f}$ along the frontal direction and $F^{2D}_{s}$ along the lateral direction $\frac{D^{\prime}}{4}$ times within Lifting module, as illustrated in Fig.~\ref{fig:network_archtecture}. Finally, the lifted 3D feature maps, $F^{3D}_{f}$ and $F^{3D}_{s}$, are concatenated channel-wise to form the combined 3D feature map $F_{B}\in{R}^{1024\times\frac{D^{\prime}}{4}\times\frac{H^{\prime}}{4}\times\frac{W^{\prime}}{4}}$.
Then, $F_{B}$ is spatially upsampled using three ResNet blocks illustrated in Fig.~\ref{fig:network_archtecture}, each followed by an Upsample Type 2 block illustrated in Fig.~\ref{fig:network_archtecture}. During the upsampling process, the channel dimension of the volumes is reduced. The combination of ResNet Block and Upsample Type 2 block can upsample the spatial size and preserve information from the previous layer.
And every Resnet Block outputs $f_{u_{21}}\in{R}^{256\times\frac{D^{\prime}}{4}\times\frac{H^{\prime}}{4}\times\frac{W^{\prime}}{4}}$, $f_{u_{22}}\in{R}^{128\times\frac{D^{\prime}}{2}\times\frac{H^{\prime}}{2}\times\frac{W^{\prime}}{2}}$ and $f_{u_{23}}\in{R}^{64\times D^{\prime} \times H^{\prime} \times W^{\prime}}$ from the first, second, and third blocks, respectively as shown in Fig.~\ref{fig:network_archtecture}.  $f_{u_{21}}$ and $f_{u_{22}}$ are further processed by Upsample type 1 blocks illustrated in Fig.~\ref{fig:network_archtecture}, resulting in $f_{u_{11}}\in{R}^{64\times D^{\prime} \times H^{\prime} \times W^{\prime}}$ and $f_{u12}\in{R}^{64\times D^{\prime} \times H^{\prime} \times W^{\prime}}$ to share same spatial dimension. By leveraging this process, we aim to utilize the information from multi-scale feature maps effectively.

In the second branch, we duplicate projections $I_{f}$ along the FR direction and $I_{s}$ along the LA direction by $D^{\prime}$ times to generate 3D volumes $I^{3D}_{f}$ and $I^{3D}_{s}$. These are concatenated and passed through a ResNet block, producing the feature map $F_{o}$. Finally, $F_{o}$ is concatenated with $f_{u_{11}}$, $f_{u_{12}}$, and $f_{u_{23}}$, resulting in the final feature map $F_{u}\in{R}^{224\times D^{\prime} \times W^{\prime} \times H^{\prime}}$, which provides the Reconstruction module with direct information from the original input.

For reconstruction, we apply Dropout with a probability of 0.25 to $F_{u}$ to enhance the model's generalization ability. Then, we use the U-Net++ model~\cite{zhou2018unet++} that is composed of four encoder layers, four decoder layers, and six nested dense skip connections with Deep Supervision~\cite{zhou2018unet++} to fuse and extract the 3D features from the concatenated 3D feature map $F_{u}$. The extracted 3D feature map is further passed through a U-Net to reconstruct the chest volume $R_{v}$.

In cases where the Generator (G) is trained independently without the Discriminator (D), we refer to the model as DuoLift-CNN. 

\subsection{Discriminator Architecture}
Our model's Discriminator (D) is designed to differentiate between real and fake volumes by encoding the target and reconstructed volumes into a 5x5x5 vector. This vector is used to classify the target chest volume as real and the reconstructed volume as fake. The Discriminator is composed of four convolutional layers, each with a 4x4x4 kernel, a stride of one, and padding of two, along with average pooling using a 3x3x3 kernel, a stride of two, and padding of one. This combination effectively encodes the volumes, and the Discriminator outputs the probability that the input volume belongs to either the real or fake class.

We incorporate a segmentation step that masks the target and reconstructed volumes with the lung segmentation code from LiftReg~\cite{tian2022liftreg} to encourage the model to focus on the structures inside the lung. Instead of feeding the masked volumes directly into the Discriminator, we propose calculating the similarity loss between the masked target and reconstructed volumes, which is more effective. This approach enhances the accuracy of reconstructing textural details within the lung. However, since we want DuoLift-CNN to focus on capturing large structures like the contours of the chest and lungs, we did not use masked loss during its training.

\subsection{Train}
To train DuoLift-GAN, we followed the approach outlined in X2CT-GAN, utilizing separate optimizers for the Generator (G) and Discriminator (D). The Discriminator is trained to differentiate between real target volumes and those generated by the model, while the Generator is optimized with a dual objective. First, G minimizes a similarity loss between the target and reconstructed volumes, both with and without masking. Second, G is trained to generate volumes that can successfully "fool" the Discriminator into classifying them as real, thus engaging in a zero-sum game with D. We fine-tuned hyper-parameters using the LIDC-IDRI dataset's validation set. We determined the model weights based on the best performance on the validation set. Specifically, we used a learning rate of 5e-4 for training DuoLift-CNN and 1e-4 for training DuoLift-GAN. The learning rate was reduced to one-tenth after 70 epochs and kept constant. The training was conducted with a batch size of 1, accumulating gradients over four batches before updating the model parameters. Our model was trained on one NVIDIA RTX A6000 GPU, each with 48GB of memory, for 100 epochs.

\subsection{Loss}
The loss function for our model is divided into two main components: the generator loss (G loss) and the discriminator loss (D loss). The G loss is primarily focused on making the reconstructed volume as similar as possible to the target volume and fooling the discriminator into classifying the reconstructed volume as real.

The G loss is calculated by combining the Mean Squared Error (MSE) and L1 loss between the reconstructed volume $\hat{I}$ and the target volume $I$ as:
\begin{equation}
    \mathcal{L}_{recon}^{MSE} = \frac{1}{|\Omega|}\sum_{\Omega}(I - \hat{I})^2,
\end{equation}

\begin{equation}
    \mathcal{L}_{recon}^{L1} = \frac{1}{|\Omega|}\sum_{\Omega}||I - \hat{I}||_1,
\end{equation}

In addition, the G loss includes the MSE and L1 loss between the masked reconstructed volume $\hat{MI}$ and the masked target volume $MI$, which focuses on the lung region:
\begin{equation}
    \mathcal{L}_{inside}^{MSE} = \frac{1}{|\Omega|}\sum_{\Omega}(MI - \hat{MI})^2,
\end{equation}

\begin{equation}
    \mathcal{L}_{inside}^{L1} = \frac{1}{|\Omega|}\sum_{\Omega}||MI - \hat{MI}||_1,
\end{equation}

To further refine the generator, an adversarial loss component $\mathcal{L}_{ag}$ is included to train the generator to produce volumes that can fool the discriminator. On the other hand, the D loss is designed to train the discriminator to distinguish between real and fake volumes. It is calculated using the adversarial loss $\mathcal{L}_{ad}$ :
\begin{equation}
\mathcal{L}_{ag} = - \mathbb{E}_{x \sim p_z(x)} \left[\log \left(D(G(x)) \right) \right]
\end{equation}

\begin{equation}
\mathcal{L}_{ad} = -\mathbb{E}_{y \sim p(CT)} \left[ \log D(y) \right] - \mathbb{E}_{x \sim p(x)} \left[ \log \left( 1 - D(G(x)) \right) \right]
\end{equation}
where $y \sim p(CT)$ represents samples from the real data distribution.$x \sim p(x)$ represents $x$ is X-rays from the data distribution $p(x)$ with $G(x)$ represents the generated 3D volume and $D(y)$ the discriminator's output for real data $y$.

The overall G loss is then summarized as:
\begin{equation}
    \mathcal{L}_{G} = \alpha(\mathcal{L}_{recon}^{MSE} + \mathcal{L}_{recon}^{L1}+ \mathcal{L}_{inside}^{MSE} + \mathcal{L}_{inside}^{L1}) + \beta\mathcal{L}_{ag}
\end{equation}

 The D loss is expressed as:
\begin{equation}
    \mathcal{L}_{D} = \beta\mathcal{L}_{ad}
\end{equation}
In our experiments, the hyper-parameters $\alpha$ and $\beta$ are set to 1.0 and 0.01, respectively.

\begin{table*}[t]
    \centering
    \resizebox{\textwidth}{!}{
    \begin{tabular}{lccccc} \toprule
         \multirow{2}{*}{Methods} & \multicolumn{5}{c}{LIDC-IDRI} \\ \cmidrule{2-6}
         & \multicolumn{5}{c}{Single-view} \\ \cmidrule{2-6}
         & PSNR(2D) $\uparrow$ & PSNR(3D) $\uparrow$ & SSIM(2D)$\uparrow$  & SSIM(3D) $\uparrow$ & LPIPS(VGG)$\downarrow$ \\
         X2CT-GAN & $22.225\pm3.430$ & $17.906\pm1.684$ & $0.461\pm0.085$ & $0.512\pm0.078$ & $0.342\pm0.054$ \\
         X2CT-CNN & $23.203\pm3.675$ & $18.725\pm1.887$ & $0.533\pm0.086$ & $0.580\pm0.079$ & $0.381\pm0.056$\\
         DuoLift-GAN & $21.572\pm2.713$ & $18.253\pm1.558$ & $0.530\pm0.071$ & $0.507\pm0.077$ & \pmb{$0.331\pm0.053$} \\
         DuoLift-CNN & \pmb{$23.779\pm3.447$} & \pmb{$19.200\pm1.782$} & \pmb{$0.599\pm0.072$} & $0.571\pm0.077$ & $0.371\pm0.055$ \\ \cmidrule(lr){2-6}
         \multirow{2}{*}{ } & \multicolumn{5}{c}{Double-view} \\ \cmidrule(lr){2-6}
         X2CT-GAN & $25.727\pm3.169$ & $21.577\pm1.109$ & $0.622\pm0.064$ & $0.655\pm0.058$ & $0.275\pm0.047$ \\
         X2CT-CNN & $27.406\pm3.524$ & $22.752\pm1.283$ & $0.689\pm0.059$ & \pmb{$0.719\pm0.054$} & $0.313\pm0.053$ \\
         DuoLift-GAN & $26.695\pm3.462$ & $22.006\pm1.200$ & $0.676\pm0.057$ & $0.663\pm0.062$ & \pmb{$0.249\pm0.047$} \\
         DuoLift-CNN & \pmb{$27.850\pm3.666$} & \pmb{$23.043\pm1.428$} & \pmb{$0.727\pm0.056$} & $0.712\pm0.066$ & $0.276\pm0.051$ \\ \bottomrule
    \end{tabular}}
    \caption{Evaluation results (Mean ± Standard Deviation) of metrics with SSIM(2D) and PSNR(2D) on the LIDC-IDRI dataset with single and double view settings on our model and X2CT\cite{ying2019x2ct}.}
    \label{tab:2d3d_result_combined}
\end{table*}

\section{Experiments}
\subsection{Dataset Description}
\paragraph{LIDC-IDRI} The LIDC-IDRI dataset \cite{Armato2011} contains thoracic CT scans used for diagnostic and lung cancer screening, with annotations identifying and delineating lung lesions. We utilized the preprocessed data from X2CT-GAN\cite{ying2019x2ct}, which includes chest volume projections processed through CycleGAN~\cite{zhu2017unpaired} to bridge the gap between projected images and actual X-rays. The original dataset comprises a training set and a test set, containing 916 and 102 CT images, respectively. We further divided the training set into 816 images for training and 100 images for validation. 
The CT images are linearly interpolated to a resolution of 128×128×128, and projections are simulated using the official script from \cite{ying2019x2ct}, which generates projections with a resolution of 128×128.

\subsection{Evaluation Metrics}
\paragraph{Structural Similarity Index Measure (SSIM)}The Structural Similarity Index Measure (SSIM)\cite{wang2004image} assesses image quality by quantifying structural similarities. This study utilizes both 2D and 3D versions of SSIM to compare the structural similarity between reconstructed 3D chest volumes and target CT volumes. SSIM(2D) evaluates the volume slice by slice using a 2D Gaussian kernel and then averaging the values across all slices, while SSIM(3D) applies a 3D Gaussian kernel to the entire volume \cite{SSIM3Dgithub}. A higher SSIM score indicates greater structural similarity between the volumes.

\paragraph{Peak Signal-to-Noise Ratio (PSNR)}Peak Signal-to-Noise Ratio (PSNR) measures image quality by comparing the maximum signal value to the noise level expressed in decibels (dB). There are 2D and 3D versions of PSNR. PSNR(2D) applies PSNR on each slice of volume, and PSNR(3D) applies PSNR to the entire volume. 

\paragraph{Learned Perceptual Image Patch Similarity (LPIPS)}Learned Perceptual Image Patch Similarity (LPIPS)\cite{zhang2018perceptual} measures perceptual similarity between images using pre-trained networks, closely aligning with human visual perception. In this study, we use LPIPS with the VGG network to evaluate reconstruction quality, where a lower LPIPS score indicates higher visual quality. To evaluate the visual quality of the entire 3D structure using LPIPS, we compute the metric for each slice and then take the mean value across all slices in the reconstructed volume.

\paragraph{Dice similarity coefficient (DICE)}The Dice Similarity Coefficient (DICE) is a statistical measure used to assess similarity between two samples. It is calculated between the segmentation masks of reconstructed and target volumes to evaluate the similarity of their regions of interest (ROIs). A higher DSC score indicates greater similarity between the target and reconstructed volumes.

\subsection{CT Reconstruction from XRays}
We compare DuoLift-CNN and DuoLift-GAN with X2CT-CNN and X2CT-GAN, publicly available state-of-the-art (SOTA) models for single-view and double-view chest CT reconstruction. To ensure a fair comparison, we use the official implementation and hyperparameters from the X2CT repository\footnote{https://github.com/kylekma/X2CT} and train the X2CT models using the same dataset split as for DuoLift-CNN and DuoLift-GAN. As shown in Table~\ref{tab:2d3d_result_combined}, DuoLift-CNN and DuoLift-GAN consistently outperform X2CT-CNN and X2CT-GAN in terms of PSNR, SSIM, and LPIPS on the LIDC-IDRI dataset. This trend is observed across both single-view and double-view settings, demonstrating that the DuoLift-variants framework achieves better reconstruction quality than existing SOTA models. 

\begin{figure}[ht]
    \centering
    \includegraphics[width=0.35\textwidth]{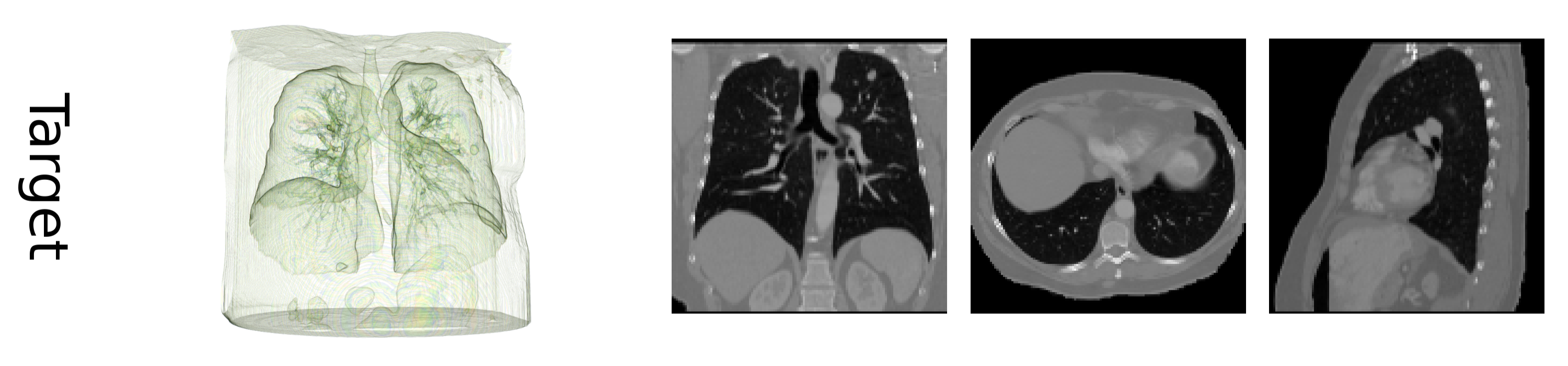}
    \includegraphics[width=0.35\textwidth]{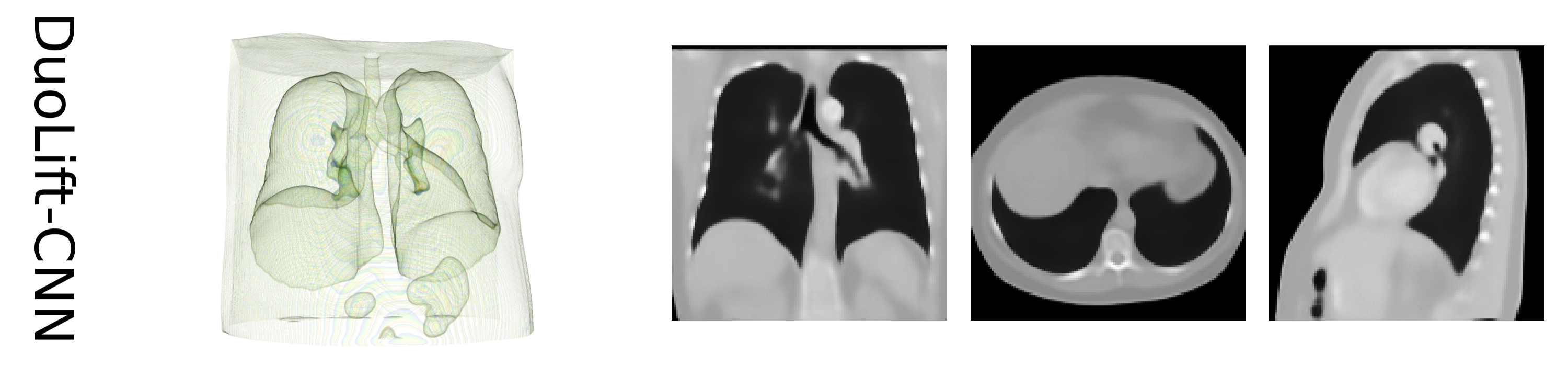}
    \includegraphics[width=0.35\textwidth]{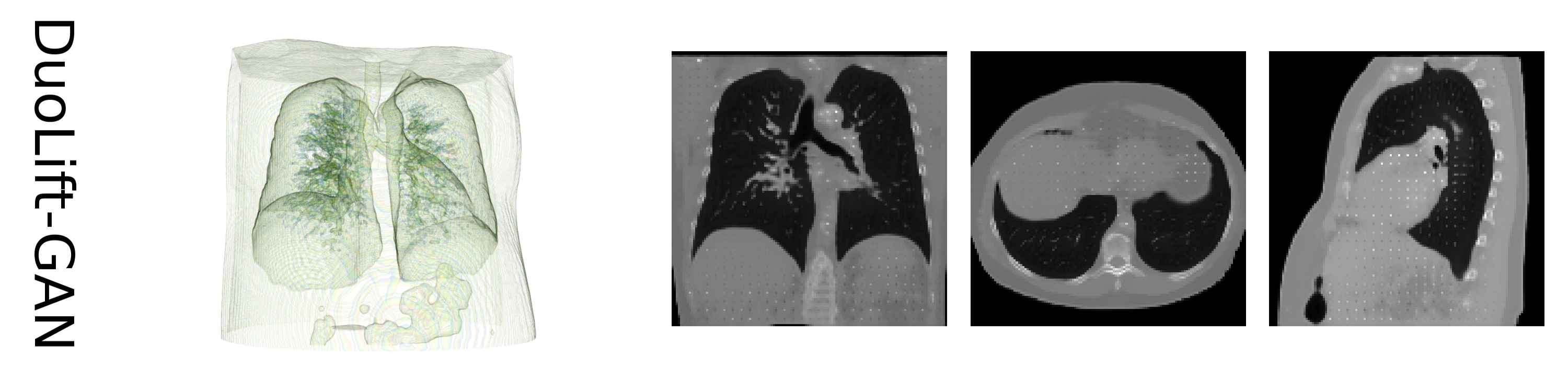}
    \includegraphics[width=0.35\textwidth]{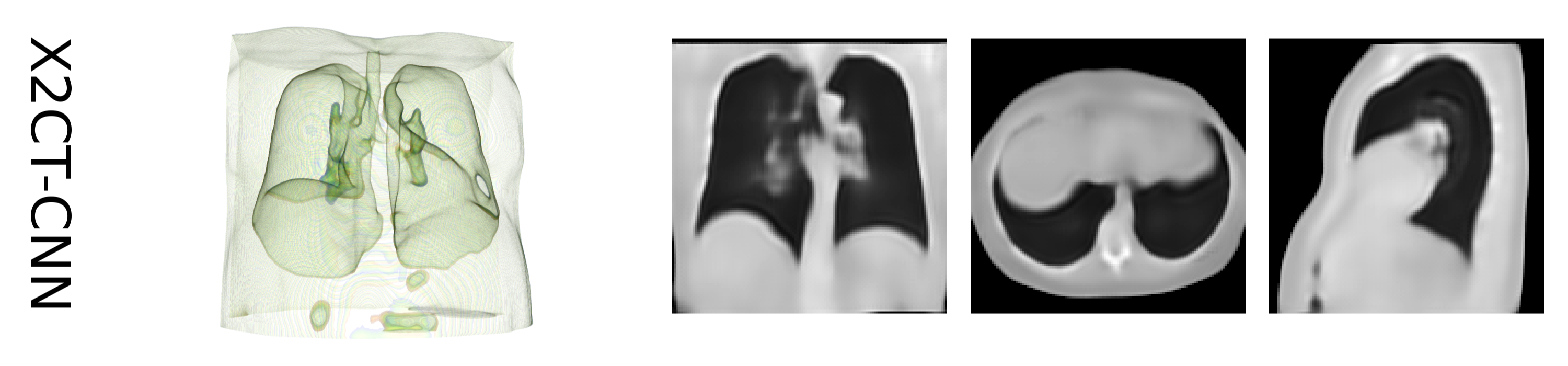}
     \includegraphics[width=0.35\textwidth]{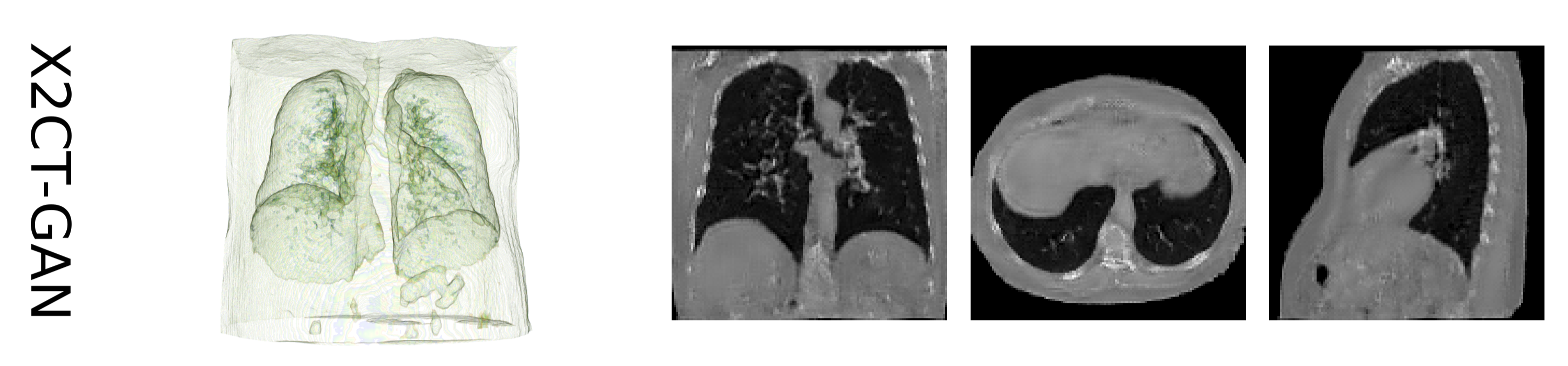}
    \caption{The visualization of reconstructed volumes from the chosen sample in the test set of the LIDC-IDRI dataset, grouped by the corresponding model. The left side shows 3D renderings, while the right side presents the reconstructed volume's coronal, axial, and sagittal slices.}
    \label{fig:lidc_recon_visual}
\end{figure}

Additionally, we observe discrepancies between DuoLift-CNN and DuoLift-GAN when evaluating reconstruction quality with SSIM and LPIPS(VGG). We will conduct further qualitative and quantitative shape analyses in the following sections to understand these differences.
\paragraph{Why does DuoLift-GAN perform better on LPIPS(VGG) but worse on SSIM?} 
We analyzed a randomly selected test sample from the LIDC-IDRI dataset to investigate metric inconsistencies by reconstructing volumes using DuoLift-GAN, DuoLift-CNN, X2CT-GAN, and X2CT-CNN. Quantitative and qualitative results are shown in Table~\ref{tab:5x5_table_method_compare} and Figure~\ref{fig:lidc_recon_visual}, respectively. DuoLift-CNN achieves higher SSIM and PSNR scores than DuoLift-GAN, consistent with the test set findings in Table~\ref{tab:2d3d_result_combined}. However, DuoLift-GAN provides more detailed lung structures, aligning with LPIPS and highlighting its advantage in capturing internal features. This indicates that SSIM and PSNR emphasize larger structures, whereas LPIPS more effectively assesses finer lung details.
\begin{table}[h]
    \centering
    \scriptsize 
    \setlength{\tabcolsep}{2pt} 
    \renewcommand{\arraystretch}{1} 
    \begin{tabular}{lccccc} 
        \toprule
        Methods & PSNR(2D)$\uparrow$  & PSNR(3D)$\uparrow$  & SSIM(2D)$\uparrow$  & SSIM(3D)$\uparrow$  & LPIPS(VGG)$\downarrow$  \\ 
        \midrule
        1.DuoLift-GAN& $23.471$ & $21.504$ & $0.678$ & $0.644$ & \pmb{$0.251$} \\ 
        2.DuoLift-CNN& \pmb{$24.363$} & \pmb{$22.400$} & \pmb{$0.725$} & $0.697$ & $0.273$ \\ 
        3.X2CT-GAN & $22.906$ & $20.908$ & $0.604$ & $0.635$ & $0.298$ \\ 
        4.X2CT-CNN & $23.749$ & $21.730$ & $0.674$ & \pmb{$0.702$} & $0.314$ \\ 
        \bottomrule
    \end{tabular}
    \caption{The quantitative result of the reconstructed volume shown in Figure~\ref{fig:lidc_recon_visual}.}
    \label{tab:5x5_table_method_compare}
\end{table}

Interestingly, although X2CT-GAN provides more detailed reconstructions, its LPIPS score is higher, indicating \emph{lower} perceptual similarity to the target volume compared to DuoLift-CNN. This is likely due to X2CT-GAN’s tendency to produce discontinuous structures, as illustrated in Figure~\ref{fig:lidc_recon_visual}, which LPIPS interprets as less similar to the continuous structures in the target volume. Consequently, despite X2CT-GAN's detailed reconstructions, its LPIPS score is lower compared to DuoLift-CNN and DuoLift-GAN.


\paragraph{How does DuoLift-GAN perform reconstructing ROIs?}
We calculate the DICE coefficient for the region of interest (ROI) to evaluate reconstruction quality further. We use pre-trained medical image segmentation models, TotalSegmentator~\cite{d2024totalsegmentator} and Lungmask~\cite{wang2009automated}\footnote{github: https://github.com/JoHof/lungmask}, to obtain vessel and lung segmentation masks from both the reconstructed and target volumes. The DICE coefficient is computed for the lung and vessel segmentation masks, comparing the reconstructed volumes to the target volumes on the LIDC-IDRI dataset, as detailed in Table~\ref{tab:2d3d_result_DICE}.

\begin{table}[htp]
    \centering
    \setlength{\tabcolsep}{2.0pt} 
    \renewcommand{\arraystretch}{1} 
    \begin{tabular}{lcc} \toprule
         \multirow{2}{*}{Methods} & \multicolumn{2}{c}{LIDC-IDRI} \\ \cmidrule{2-3}
         &DICE(LungSeg)$\uparrow$&DICE(VesselSeg)$\uparrow$\\  \cmidrule{2-3}
         X2CT-GAN & \pmb{$0.931\pm0.026$} & $0.097\pm0.048$\\
         X2CT-CNN & $0.903\pm0.039$ & $0.008\pm0.010$\\
         DuoLift-GAN & $0.928\pm0.029$ & \pmb{$0.098\pm0.050$} \\
         DuoLift-CNN & \pmb{$0.931\pm0.028$} & $0.058\pm 0.040$  \\ \bottomrule
    \end{tabular}
    \caption{ The \textbf{DICE$\uparrow$} (Mean ± Standard Deviation) of the lung and vessel segmentation masks between the reconstructed and target volumes of the LIDC-IDRI test set. The volumes are reconstructed in the double-view setting.}
    \label{tab:2d3d_result_DICE}
\end{table}

According to the results presented in Table~\ref{tab:2d3d_result_DICE}, DuoLift-CNN achieves the highest DICE score for lung masks, indicating its superior ability to reconstruct lung regions when comparing the reconstructed and target volumes. This outcome aligns with the design of DuoLift-CNN, which is optimized for capturing large structural shapes.

Furthermore, DuoLift-GAN achieved the best reconstruction performance for vessels in the double-view setting. This finding aligns with the qualitative analysis in Figure~\ref{fig:lidc_recon_visual}, where DuoLift-GAN demonstrated superior capability in reconstructing the finer details within the lung compared to DuoLift-CNN.



\paragraph{2D and 3D SSIM Comparison} Quantitative results in Table~\ref{tab:2d3d_result_combined} show a significant difference in SSIM(2D) scores between X2CT-CNN and DuoLift-CNN, while the difference in SSIM(3D) scores is much smaller. As illustrated in Figure~\ref{fig:lidc_recon_visual}, DuoLift-CNN's reconstructed volume exhibits more precise and clear edges than X2CT-CNN, making its contour more similar to the target volume. This difference is due to SSIM(2D) evaluating volumes slice by slice, accumulating pixel-wise differences, and amplifying the gap between the target and reconstructed volumes. Conversely, SSIM(3D) assesses the entire volume as a whole 3D structure, which mitigates the cumulative effect of pixel-wise differences.

\subsection{Ablation Study}
In this section, we conduct ablation studies on a private dataset with 699 volumes for training and 100 volumes for testing to answer the following two questions: 1) How does the lifting module affect the reconstruction quality? and 2) what is the most effective way to utilize the masked chest volume?

\paragraph{Do we need the duo lifting module?}
We trained DuoLift-CNN on the private dataset in the double-view reconstruction setting to assess the impact of the Lifting module shown in Fig.~\ref{fig:network_archtecture}. The results presented in Table~\ref{tab:Structual_Ablation} show that introducing the dual lift branches into the model led to improved PSNR, SSIM, LPIPS, and DICE scores for the reconstruction task. 

For comparison, we evaluated two alternative configurations. The first configuration uses dual branches but replaces the duplication operation in the Lifting module with a 2D convolution layer in both branches, whose result is shown in the second row in Table~\ref{tab:Structual_Ablation}. In the second configuration, we keep the 2nd branch and show the result in the third row in Table~\ref{tab:Structual_Ablation}. As shown in Table~\ref{tab:Structual_Ablation}, DuoLift-CNN, combining both the Lifting module and dual branches, achieved the best performance.


\begin{table}[htp]
    \centering
    \scriptsize 
    \setlength{\tabcolsep}{2.5pt} 
    \renewcommand{\arraystretch}{1} 
    \begin{tabular}{cccccc} 
    \toprule
    L&DB&PSNR(2D)$\uparrow$ &SSIM(2D)$\uparrow$ &LPIPS(VGG)$\downarrow$ &DICE(LungSeg)$\uparrow$ \\
    \hline
    \hfil \cmark & \hfil \cmark & \pmb{$35.873\pm4.484$}&\pmb{$0.843\pm0.042$}&\pmb{$0.160\pm0.040$}&\pmb{$0.946\pm0.020$}\\
    \hfil \xmark & \hfil \cmark& $32.445\pm4.567$&$0.757\pm0.053$&$0.249\pm0.049$&$0.888\pm0.038$ \\
    \hfil \cmark & \hfil \xmark& $35.590\pm4.519$&$0.834\pm0.044$&$0.173\pm0.042$&$0.944\pm0.019$ \\
    \hline
    \end{tabular}
    \newline\newline
    \caption{The ablation study of the lifting module. \textbf{L} denotes lift by repeating along certain axis, and \textbf{DB} denotes duo branch, where lifting is applied to both the feature map and the original projections by repeating along certain axis. \textbf{DB} and \textbf{L} is what we adopted in our proposed method.}
    \label{tab:Structual_Ablation}
\end{table}

\paragraph{How should the masked chest volume be used?} 
This experiment explores the optimal way to utilize the reconstructed chest and masked target volumes based on an ablation study in a double-view setting. We refer to the DuoLift-GAN without $\mathcal{L}_{inside}^{MSE} + \mathcal{L}_{inside}^{L1}$ as \textbf{Ablation 1}, and the DuoLift-CNN with $\mathcal{L}_{inside}^{MSE} + \mathcal{L}_{inside}^{L1}$ as \textbf{Ablation 2}. Furthermore, instead of computing $\mathcal{L}_{inside}^{MSE} + \mathcal{L}_{inside}^{L1}$, one has the option to pass the masked reconstructed and target volumes to the Discriminator. This is denoted as \textbf{Ablation 3} in Table~\ref{tab:Masked_Ablation}.

\begin{table}[htp]
    \centering
    \scriptsize 
    \setlength{\tabcolsep}{2pt} 
    \renewcommand{\arraystretch}{1} 
    \resizebox{\linewidth}{!}{\begin{tabular}{ccccccccc} 
        \toprule
    &G &D &SL&DI&PSNR(2D)$\uparrow$&SSIM(2D)$\uparrow$&LPIPS(VGG)$\downarrow$&DICE(VesselSeg)$\uparrow$ \\
    \hline
    Ablation 1&\hfil \cmark & \cmark & \xmark & \xmark &$35.177\pm4.478$&$0.822\pm0.044$&$0.154\pm0.041$&$0.239\pm0.061$ \\
    Ablation 2&\hfil \cmark & \xmark & \cmark & \xmark &$\pmb{35.799\pm4.477}$&$\pmb{0.842\pm0.0430}$&$0.162\pm0.041$&$0.212\pm0.056$ \\
    Ablation 3&\hfil \cmark & \cmark & \xmark & \cmark &$31.946\pm3.103$&$0.765\pm0.0445$&$0.219\pm0.038$&$0.115\pm0.038$ \\
    DuoLift-GAN&\hfil \cmark & \cmark & \cmark & \xmark &$34.678\pm4.410$&$0.804\pm0.0477$&$\pmb{0.147\pm0.037}$&$\pmb{ 0.261\pm0.057}$ \\
    \hline
    \end{tabular}}
    \newline\newline
    \caption{The ablation study exploring the use of masked reconstructed and target volumes. \textbf{G} denotes the generator and \textbf{D} denotes the discriminator. \textbf{SL} refers to calculating the similarity loss between masked volumes, while \textbf{DI} involves using the masked volume as input for D.}
    \label{tab:Masked_Ablation}
\end{table}

DuoLift-GAN achieves the lowest LPIPS score and the highest DICE for vessel segmentation masks, indicating good textural detail and visual quality shown in Table~\ref{tab:Masked_Ablation}. The Ablation 2 model shows better PSNR and SSIM compared to DuoLift-GAN, which aligns with our previous understanding that CNN-based models like the Ablation 2 model can more effectively reconstruct larger shapes (e.g., the lung). Additionally, Ablation 1 displays superior SSIM to DuoLift-GAN but underperforms in terms of LPIPS and DICE(VesselSeg). This supports our rationale for the masked similarity measure, aiming to reduce undetected artifacts (e.g., disconnected structures) by the Discriminator. DuoLift-GAN outperforms the Ablation 3 model across all metrics. Furthermore, Ablation 3 exhibits worse performance than the Ablation 1 model. One hypothesis is that the Discriminator, designed for classification rather than reconstruction, focuses on global perception instead of similarity between fine-grained details. Consequently, it struggles to differentiate the reconstructed and target volume when losing context (e.g., the tissue and organs outside of the lung).

\section{Limitation}
Despite offering superior reconstruction quality compared to X2CT, our method has limitations. We have yet to incorporate advanced generative models like diffusion models, which excel in generating high-fidelity 2D images due to the high memory demands of dense 3D prediction tasks such as CT reconstruction. Additionally, instead of lifting 2D feature maps to 3D via repetition of 2D feature maps via depth axis, explicitly back-projecting 2D features based on acquisition geometry could reduce geometry bias and improve model generalization. Furthermore, while LPIPS is widely used for image quality assessment, its training on natural images may only partially capture the nuances of grayscale medical X-rays, leading to potential bias. The accuracy of DICE scores also depends on the generalization capability of the segmentation models, highlighting the need for further analysis with human-annotated segmentation masks of the lungs and vessels.

\section{Conclusion}
In this study, we introduce DuoLift-variants, a novel approach for reconstructing 3D chest volumes from single-view and double-view X-rays. By using dual branches to lift 2D images and feature maps into 3D space, DuoLift-variants effectively mitigate the challenges of mapping between 2D and 3D images. The integration of a masked loss function further enhances the reconstruction of critical regions, improving structural detail and visual quality. Our extensive evaluation of the LIDC-IDRI dataset demonstrates that DuoLift-GAN significantly outperforms existing reconstruction methods in accuracy and visual realism, highlighting its potential as a practical tool in clinical settings where CT access is limited. Additionally, we provide a comprehensive quantitative and qualitative analysis, assessing reconstruction quality from pixel, anatomical structure, and perceptual levels. Our study offers valuable insights into evaluation metrics and serves as a foundation for further chest CT reconstruction quality analysis.

\bibliography{aaai25}
\end{document}